\documentclass[11pt,a4paper]{article}
\usepackage[T1]{fontenc}
\usepackage[utf8]{inputenc}
\usepackage{lmodern}
\usepackage[margin=2.5cm]{geometry}
\usepackage{amsmath,amssymb}
\usepackage{graphicx}
\usepackage{booktabs}
\usepackage{array}
\usepackage{float}
\usepackage[font=small,labelfont=bf,labelsep=period]{caption}
\usepackage{hyperref}
\usepackage{xurl}
\usepackage{xcolor}
\hypersetup{colorlinks=true,linkcolor=blue!50!black,citecolor=blue!50!black,urlcolor=blue!50!black}
\usepackage{microtype}

\newcommand{\rf}[2]{\hyperlink{ref:#1}{#2}}
\newcommand{\bibent}[2]{\item[]\hypertarget{ref:#1}{}#2}
\newcommand{\doi}[1]{doi:\href{https://doi.org/#1}{\nolinkurl{#1}}}
\newcommand{\para}[1]{\textit{#1}}

\title{\textbf{Higher-order plasticity: from information to function}}
\author{Igor Branchi\thanks{E-mail: \href{mailto:igor.branchi@iss.it}{igor.branchi@iss.it}. ORCID: \href{https://orcid.org/0000-0003-4484-3598}{0000-0003-4484-3598}}\\[0.5em]
\small Center for Behavioral Sciences and Mental Health, Istituto Superiore di Sanit\`a\\
\small Viale Regina Elena 299, 00161 Rome, Italy}
\date{November 2026}

\begin{document}
\maketitle

\begin{abstract}
\noindent Plasticity, the capacity of a system to change, can be defined as the inverse of configurational constraint. Current network measures estimate this constraint from pairwise interactions, although complex systems may also contain higher-order interactions. Here we ask how much additional constraint higher-order interactions impose beyond what pairwise statistics capture. Addressing this question calls for a conceptual shift in the study of higher-order networks: from information to function, that is, from describing statistical dependencies among elements to understanding how their interactions give rise to system-level properties, such as the capacity for change. In simulated binary networks, we quantified constraint as the expected structural cost of changing a single element and compared each system with its pairwise maximum-entropy description. The pairwise description missed up to the entire constraint, with the discrepancy depending on the strength and architecture of higher-order interactions. This functional constraint was poorly captured by informational indices, including O-information and information beyond pairs. Accounting for higher-order interactions also improved the prediction of actual change following perturbation, which pairwise plasticity substantially overestimated in strongly constrained systems. These findings show that plasticity depends on constraints arising from both pairwise and higher-order interactions. More broadly, they indicate how a shift from an informational perspective toward a functional understanding of higher-order interactions enables the quantification of their contribution to system-level properties. The functional perspective thus complements the informational description of complex systems.
\end{abstract}

\begin{quote}
\noindent\textbf{Many complex systems, from the brain to ecological and social networks, contain interactions that involve three or more elements at once. These higher-order interactions are usually studied with informational measures, which describe how the activity of the elements is statistically related. The question explored here is how to approach higher-order interactions to measure functional properties, such as the capacity of a system to change, that is, its plasticity. Using simulated networks, we measure the constraint that interactions impose on change and compare each system with its pairwise description, a model that reproduces all its pairwise statistics. Overlooking higher-order functional interactions can substantially misestimate plasticity, and informational measures capture the missing constraint only in part. These results indicate that informational and functional approaches address different aspects of higher-order interactions. They also support the operationalization of plasticity as inverse configurational constraint, which predicts how much a system changes.}
\end{quote}

\section{Introduction}

Plasticity, the capacity of a system to change, can be defined as the inverse of the constraint that the interactions among its elements impose (\rf{branchi2022}{Branchi, 2022}, \rf{branchi2026a}{2026a}). Network measures of plasticity estimate this constraint from pairwise connectivity (\rf{dellicolli2024}{Delli Colli et al., 2024}; \rf{dellicolli2025}{Delli Colli et al., 2025}; \rf{viglione2025}{Viglione et al., 2025}; \rf{dellicolli2026}{Delli Colli et al., 2026}). Many complex systems, including the brain, also contain interactions among three or more elements (\rf{battiston2020}{Battiston et al., 2020}; \rf{luppi2022}{Luppi et al., 2022}). How much additional constraint these interactions impose beyond a pairwise description, and whether informational measures capture it, remains unclear.

Higher-order structure is usually quantified with informational measures, such as the information beyond pairs (\rf{schneidman2003}{Schneidman et al., 2003}) or O-information (\rf{rosas2019}{Rosas et al., 2019}). These measures characterize higher-order statistical dependencies, which do not necessarily reflect the underlying higher-order mechanisms (\rf{rosas2022}{Rosas et al., 2022}). In that framework, the statistical patterns produced by a system are its higher-order behavior. Here we distinguish two readings of behavior: an informational reading, which describes the statistical dependencies, and a functional reading, which describes how strongly the interactions, weighted by how much they are realized, constrain change. The four wheels of a car are redundant from an informational standpoint, because the rotation of one wheel predicts the rotation of the others. From a functional standpoint they are synergistic, because the function of moving the car binds them to move together. The same coupling represents redundancy for information and synergy for function, because the system's function depends on the joint contribution of its elements.

Here we examine higher-order interactions from a functional perspective. We ask how much constraint they impose beyond a pairwise description, and whether informational measures capture this constraint. We quantify it with the local constraint $L$. Its higher-order part is the difference in $L$ between a system and its pairwise maximum-entropy description. Using simulations, we compare this quantity with informational measures, map the conditions under which it is large, and test whether plasticity measured with and without higher-order interactions predicts how much a system actually changes after a perturbation.

\section{Model and measures}

We consider systems of $N$ binary elements, $s_i = \pm 1$, with structural cost (called energy in the Ising model)
\begin{equation}
E(s) = -\sum_i h_i s_i - \sum_{i<j} J_{ij}\, s_i s_j - \sum_{i<j<k} K_{ijk}\, s_i s_j s_k ,
\end{equation}
where each pair and each triplet is counted once. The stationary distribution is $p(s) = \exp(-E(s))/Z$, computed exactly by enumerating all configurations.

Plasticity is defined as $P = 1/C$ (\rf{branchi2026a}{Branchi, 2026a}). As operational measure of $C$ we use the local constraint $L$, the expected change in structural cost associated with flipping a randomly selected element, averaged over the stationary distribution
\begin{equation}
L = \frac{2}{N}\Big[\sum_i h_i \langle s_i\rangle + 2\sum_{i<j} J_{ij}\langle s_i s_j\rangle + 3\sum_{i<j<k} K_{ijk}\langle s_i s_j s_k\rangle\Big].
\end{equation}

Pairwise and third-order couplings contribute to $L$ according to how much they affect the average cost of changing a single element. In previous work, plasticity was quantified from network connectivity as $P = n/\sum |w_{ij}|$ (\rf{branchi2026a}{Branchi, 2026a}). Both measures quantify realized configurational constraint, through different operationalizations. The local constraint extends the network-based approach to plasticity by accounting for the realized contribution of both pairwise and higher-order interactions.

In the equilibrium model, $L$ can equivalently be written as the average Kullback--Leibler divergence, in nats, between the stationary distribution and the distribution obtained by flipping one element,
\begin{equation}
L = \frac{1}{N}\sum_i D_{\mathrm{KL}}\big(p(s)\,\|\,p(F_i s)\big),
\end{equation}
where $F_i$ inverts the state of element $i$. This identity holds by construction and is a reformulation of $L$, not an additional result. It shows that $L$ and the informational indices are expressed with the same tool, so that their comparison is made on equal terms. $L$ is therefore itself a function of the distribution, defined with respect to a specific functional transformation, the hypothetical flipping of a single element.

The higher-order contribution is $\Delta L = L(\text{system}) - L(\text{pairwise})$. The pairwise description is the maximum-entropy model with the same means and pairwise correlations as the system (\rf{jaynes1957}{Jaynes, 1957}; \rf{schneidman2003}{Schneidman et al., 2003}). It describes the system as seen by any measure based on pairwise statistics. $\Delta L$ is not simply the direct contribution of the third-order couplings. It also includes the change in the effective contribution of fields and pairwise couplings between the system and a pairwise description fitted to the same moments. It therefore measures the overall functional difference from the pairwise reference. We compared $\Delta L$ with two informational measures: the information beyond pairs, that is, the Kullback--Leibler divergence between system and pairwise description (\rf{schneidman2003}{Schneidman et al., 2003}), and O-information (\rf{rosas2019}{Rosas et al., 2019}).

\section{Methods}

All analyses used simulated systems only (Table~\ref{tab:sims}). For each simulated system, the stationary distribution and the corresponding functional and informational measures were computed by exact enumeration. The pairwise maximum-entropy description was fitted numerically using L-BFGS followed by Newton refinement, with a maximum error in the matched means and correlations below $10^{-7}$.

\begin{table}[H]
\centering
\caption{Overview of the simulations.}
\label{tab:sims}
\small
\begin{tabular}{@{}>{\raggedright\arraybackslash}p{3.4cm}>{\raggedright\arraybackslash}p{1.6cm}>{\raggedright\arraybackslash}p{3.2cm}>{\raggedright\arraybackslash}p{5.6cm}@{}}
\toprule
\textbf{Simulation} & \textbf{Elements} & \textbf{Systems} & \textbf{What varies}\\
\midrule
Parity & 3 & 5 & Third-order coupling $K = 0.25$ to 4\\
Random ensemble & 6 & 960 (768 with triplets, 192 controls) & Pairwise SD, triplet strength, number of triplets, sign mode\\
Controlled series & 6 & 160 ($20\times 8$) & Triplet strength only, architecture fixed\\
Map of conditions & 5, 6, 8 & 2,160 & Pairwise SD, triplet strength, number of triplets, sign mode\\
Change after a perturbation & 6 & 480 & Pairwise SD, triplet strength 0 to 4, number of triplets\\
\bottomrule
\end{tabular}
\end{table}

\para{Parity.} Three elements were coupled only by a third-order coupling $K$, from 0.25 to 4. All means and pairwise correlations are zero, so the pairwise description is the uniform distribution.

\para{Random ensemble.} We generated 960 systems of six elements. Fields had SD 0.1 and pairwise couplings SD 0.1, 0.3 or 0.6. Third-order couplings were placed on 1, 3, 6 or 10 random triplets. Their nominal strength was 0.25, 0.5, 1.0 or 1.5, multiplied for each triplet by a random factor uniformly distributed between 0.5 and 1.5, with aligned or mixed signs. 192 systems without triplets served as controls. As consistency checks, we compared $L$ with the analytic expression, with $\sum|J|/N$ in pairwise-only systems, and with the acceptance rate of single-element heat-bath updates (\rf{glauber1963}{Glauber, 1963}). The last relation is expected by construction.

\para{Controlled architecture series.} In 20 six-element systems, fields, pairwise couplings and the set and signs of four triplets were fixed. Only the strength of the triplets varied, over eight values from 0.1 to 1.5.

\para{Map of conditions.} We generated 2,160 systems of five, six or eight elements, crossing pairwise strength, third-order strength, number of triplets (2, 6 or 12, with at most 10 in five-element systems, which contain only 10 distinct triplets) and sign mode. Of these, 540 systems without third-order couplings ($K = 0$) served as controls. The outcomes were $\Delta L$ and the share of $L$ missed by the pairwise description, $1 - L(\text{pairwise})/L(\text{system})$.

\para{Change after a perturbation.} To test whether plasticity predicts actual change, we generated 480 six-element systems with pairwise SD 0.1 or 0.3 and 3 or 6 triplets of nominal strength 0, 0.5, 1, 2, 3 or 4 (mixed signs, 20 systems per condition). Starting from the stationary distribution, each system and its pairwise description received the same constant push, an external field of strength 0.5 toward a random pattern of the elements. The system then evolved by single-element heat-bath updates (\rf{glauber1963}{Glauber, 1963}). System change was the shift of the system toward the pushed pattern, measured as the change in the mean overlap with that pattern. It was computed exactly from the transition matrix, from 1 to 100 updates per element. This measure is defined independently of $L$. To obtain the change expected from plasticity, at each time point we estimated the relation between plasticity and change by locally weighted regression in the 480 pairwise descriptions, where the two measures of plasticity coincide, rather than in an independent sample. We then applied this relation to the systems with triplets, once using plasticity with pairwise and triplet couplings and once using plasticity measured pairwise only. Prediction errors were compared with the Wilcoxon signed-rank test in the 400 systems with nonzero third-order couplings, excluding the 80 pairwise-only systems in which the two plasticity measures coincide.

\para{Statistics.} Associations were measured with Spearman correlations. Simulations were run in Python (NumPy, SciPy, pandas, statsmodels).

\section{Results}

\begin{figure}[H]
\centering
\includegraphics[width=\textwidth]{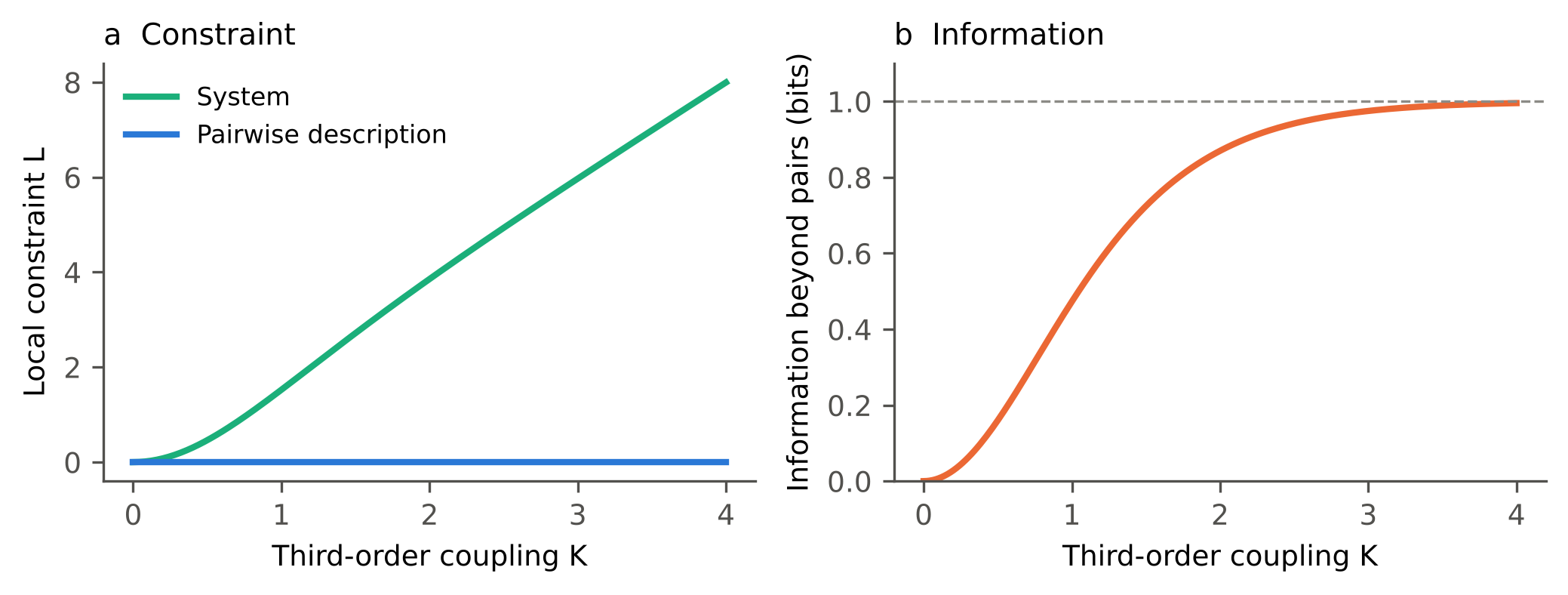}
\caption{Parity system with three elements. (a) Local constraint $L$ of the system and of its pairwise description as a function of the third-order coupling $K$. The pairwise description has no constraint. (b) Information beyond pairs, which saturates at 1 bit.}
\label{fig:1}
\end{figure}

\para{A constraint invisible to pairwise statistics.} In the parity system the pairwise description is the uniform distribution and has no constraint, $L = 0$, for any $K$. The local constraint of the system is $L = 2K\tanh K$, which grows from 0.12 at $K = 0.25$ to 8.0 at $K = 4$ (Figure~\ref{fig:1}a). The whole constraint is therefore missed by pairwise statistics. The information beyond pairs saturates at 1 bit (Figure~\ref{fig:1}b), so beyond $K \approx 2$ the system becomes much more constrained while its information beyond pairs hardly changes.

\para{Consistency checks.} The analytic expression of $L$ matched the direct average of the changes in structural cost (largest error $1\times 10^{-14}$). In the 192 systems without triplets, $|\Delta L|$ was at most $10^{-7}$. In pairwise-only systems, $L$ was closely related to $\sum|J|/N$ ($\rho = 0.97$). $1/L$ and the heat-bath acceptance were strongly related ($\rho = 0.99$), as expected by construction. Across the random ensemble, the map of conditions and the perturbation simulations, the pairwise description underestimated $L$ in 2,782 of 2,788 systems with third-order interactions and overestimated it in six. No exception occurred in the controlled series.

\begin{figure}[H]
\centering
\includegraphics[width=\textwidth]{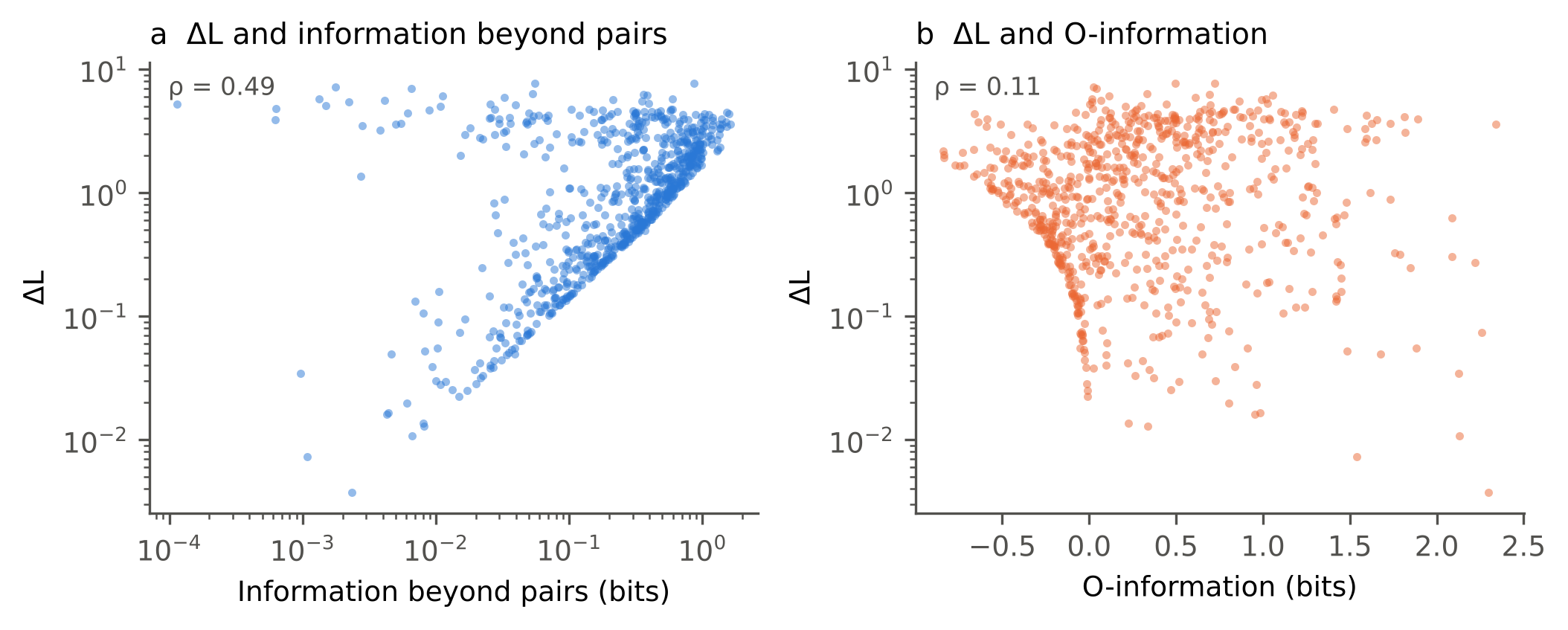}
\caption{Higher-order constraint and informational measures in 768 random systems of six elements. (a) $\Delta L$ against the information beyond pairs, both on logarithmic scales. (b) $\Delta L$ against O-information. $\rho$, Spearman correlation. One system with $\Delta L < 0$ is not shown on the logarithmic scale. Correlations include all 768 systems.}
\label{fig:2}
\end{figure}

\para{Dissociation from informational measures.} Across 768 random systems with triplets, the pairwise description missed a median 41\% of the constraint (5th--95th percentile 4--85\%). $\Delta L$ was almost unrelated to O-information ($\rho = 0.11$) and only moderately related to the information beyond pairs ($\rho = 0.49$, Figure~\ref{fig:2}). In the lowest decile of information beyond pairs, $\Delta L$ ranged from 0.021 to 5.1 between the 10th and 90th percentiles. 1,933 pairs of systems had information beyond pairs within 5\% but $\Delta L$ differing by more than threefold. These pairs involved 596 of the 768 systems and are not independent.

\begin{figure}[H]
\centering
\includegraphics[width=\textwidth]{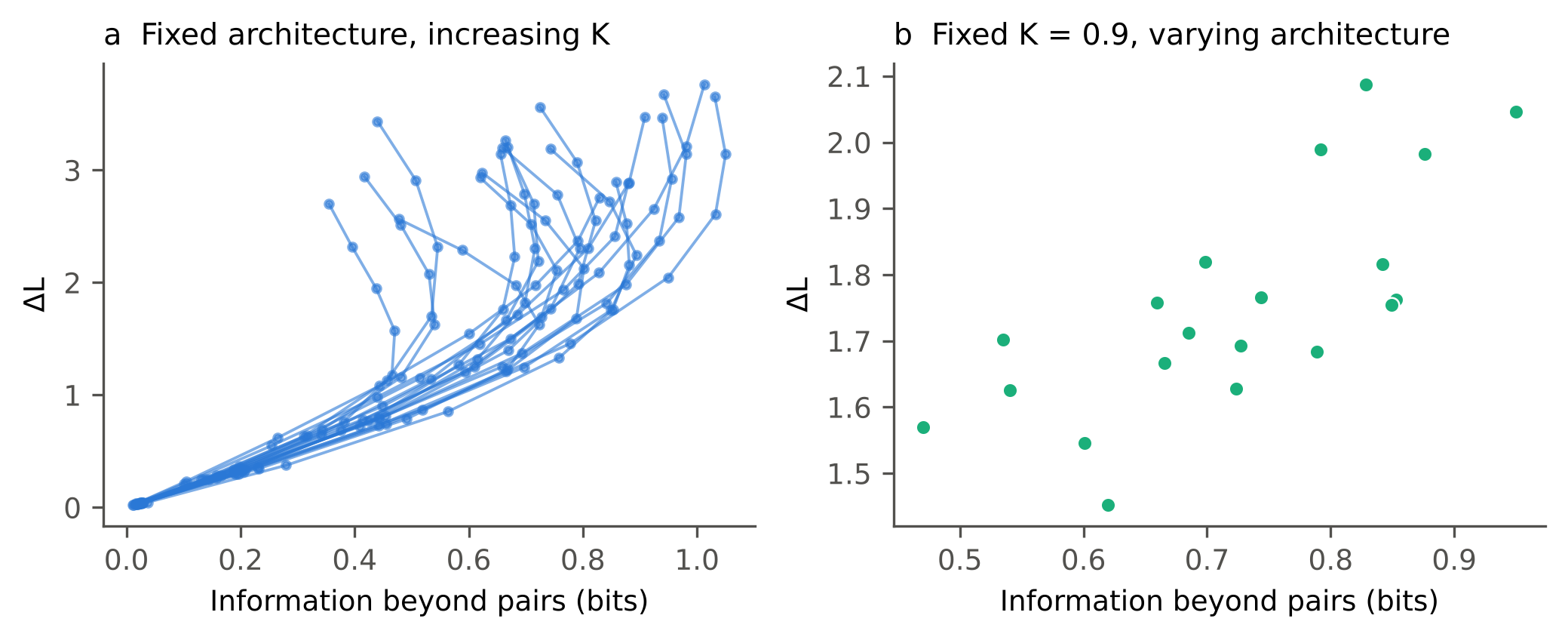}
\caption{Controlled architecture series. (a) Each line is one of 20 architectures in which fields, pairwise couplings and the set and signs of four triplets are fixed, and only the strength of the triplets increases. (b) The 20 architectures at the same triplet strength, $K = 0.9$.}
\label{fig:3}
\end{figure}

\para{Role of architecture.} When the architecture was fixed and only the strength of the triplets varied, $\Delta L$ and the information beyond pairs increased together (median within-architecture $\rho = 0.83$, Figure~\ref{fig:3}a). At equal strength across 20 architectures, $\Delta L$ ranged from 1.45 to 2.09 and information from 0.47 to 0.95 bits (Figure~\ref{fig:3}b). The relationship between informational and functional measures therefore depends on system architecture. The strength of the association within an architecture also depends on the range of triplet strengths considered. As in the parity system, the information beyond pairs tends to saturate at strong couplings while $\Delta L$ keeps increasing, and several trajectories bend backwards at the highest strengths (Figure~\ref{fig:3}a).

\begin{figure}[H]
\centering
\includegraphics[width=\textwidth]{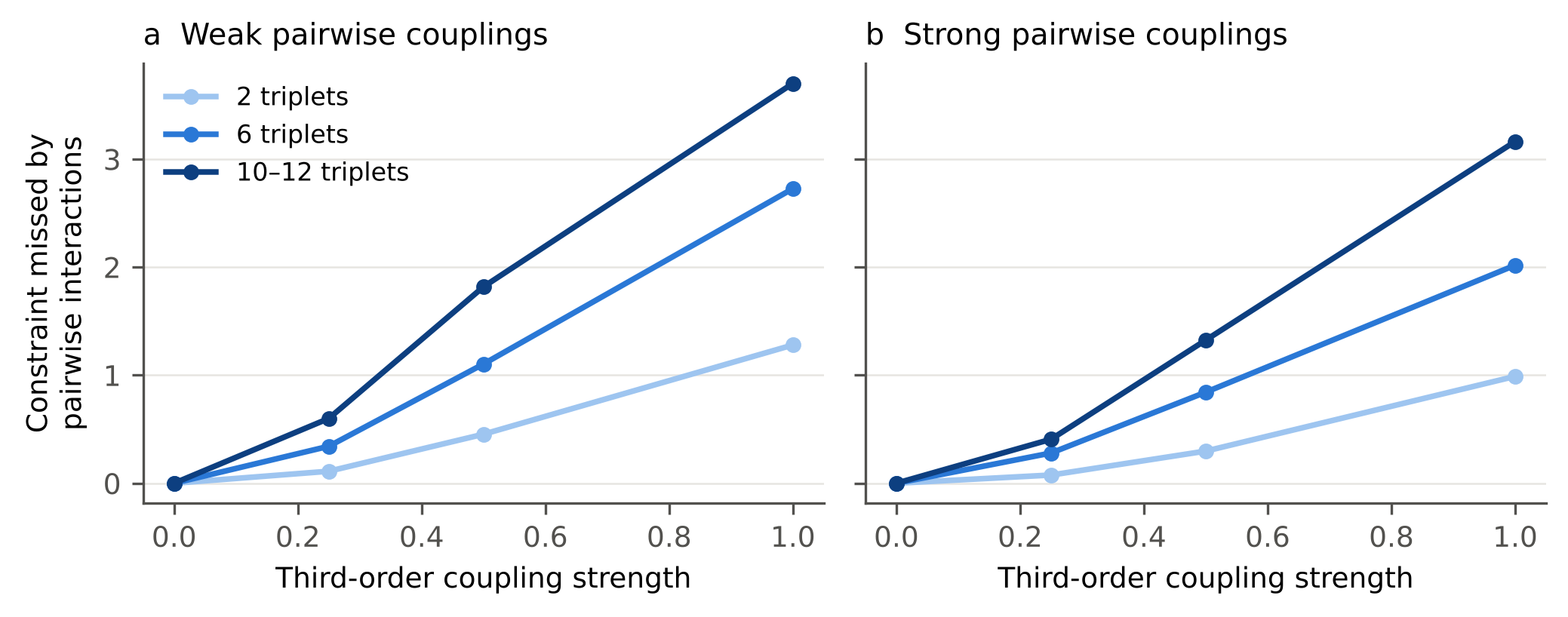}
\caption{Constraint missed by the pairwise description, $\Delta L$, as a function of third-order coupling strength, for 2, 6 and 10--12 triplets (10 in five-element systems), with weak (a) and strong (b) pairwise couplings, drawn from normal distributions with SD 0.2 and 0.5. Medians over 2,160 systems of five, six and eight elements and over sign modes.}
\label{fig:4}
\end{figure}

\para{How much constraint the pairwise description misses.} Across 2,160 systems, the constraint missed by the pairwise description increased with both the strength and the number of triplets (Figure~\ref{fig:4}). With third-order strength 1, the median $\Delta L$ was 1.28 with two triplets and 3.70 with 10--12 triplets when pairwise couplings were weak, and 0.99 and 3.16 when they were stronger. Expressed as a share of the constraint of the system, the pairwise description missed from 2\% to 72\% across conditions, more when pairwise couplings were weak. The share decreased slightly with system size (43\%, 36\% and 31\% for five, six and eight elements) and did not depend on sign mode (aligned 36\%, mixed 39\%). Without triplets, $|\Delta L|$ was at most $10^{-7}$.

\begin{figure}[H]
\centering
\includegraphics[width=0.7\textwidth]{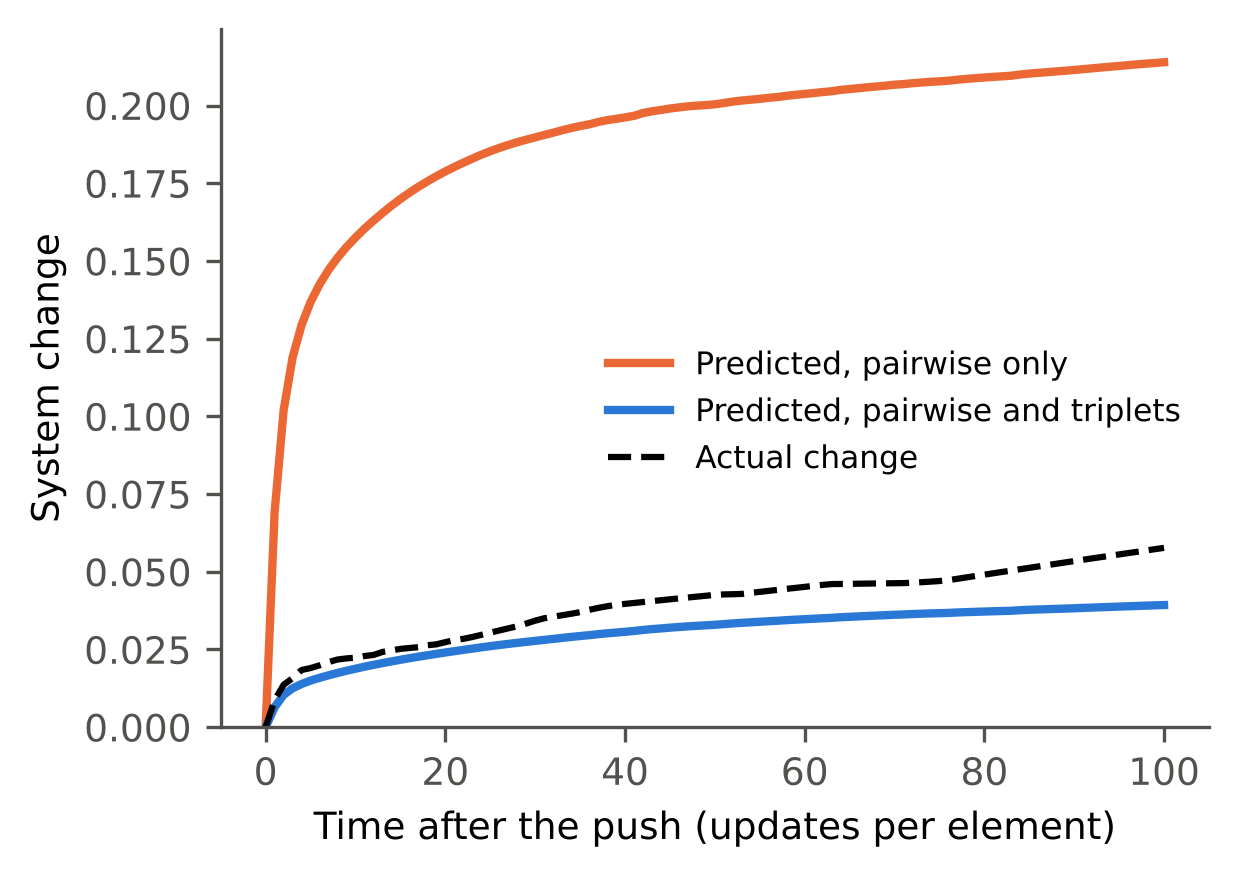}
\caption{Change expected from plasticity and actual change after a perturbation. Median over 240 six-element systems with third-order coupling $K \geq 2$. Each system received a constant push toward a random pattern (field strength 0.5) and evolved by single-element heat-bath updates. Dashed line, actual change of the system. Colored lines, change expected from plasticity measured with pairwise and triplet couplings (blue) or pairwise only (orange). The relation between plasticity and change was estimated in pairwise-only systems, where the two measures coincide.}
\label{fig:5}
\end{figure}

\para{Pairwise plasticity overestimates actual change.} In systems with strong triplets ($K \geq 2$), the change expected from plasticity measured with pairwise and triplet couplings closely followed the actual change of the system over time, with a slight underestimation (Figure~\ref{fig:5}). Plasticity measured pairwise only predicted much more change than occurred. After 20 updates per element, the median actual change was 0.027. The change expected with triplets was 0.024, and the change expected pairwise only was 0.18. After 100 updates the values were 0.058, 0.039 and 0.21. Across the 400 systems with triplets, the median absolute prediction error after 20 updates per element was 0.033 with plasticity including triplets and 0.090 with pairwise plasticity. The error was smaller with triplets in 65\% of systems (Wilcoxon signed-rank test, $p < 10^{-13}$). In systems with $K \geq 2$ the errors were 0.020 and 0.099, and the error was smaller with triplets in 79\% of systems. Across all 480 systems, plasticity with pairwise and triplet couplings was more strongly related to the actual change than plasticity measured pairwise only ($\rho = 0.92$ against 0.85 after 20 updates). With weak triplets the discrepancy was smaller and decreased over time. For $K = 1$, the pairwise description moved 77\% more than the system after 5 updates and 5\% more after 100.

\section{Discussion}

Our results show that plasticity, the capacity of a system to change, depends not only on pairwise interactions but also on higher-order interactions. The local constraint assessed here provides a functional measure of this dependence, consistent with the definition of plasticity as the inverse of configurational constraint (\rf{branchi2023}{Branchi, 2023}, \rf{branchi2026b}{2026b}).

These results also indicate that plasticity is determined by constraint itself, not by correlation or by a specific mathematical formulation. The pairwise description reproduces all means and pairwise correlations of the system, yet differs in constraint and in how much it changes after a perturbation. Whether constraint arises from pairwise or higher-order interactions, it is constraint that limits the capacity for change (\rf{branchi2026b}{Branchi, 2026b}).

The perturbation simulations provide a test of plasticity as inverse of configurational constraint. Plasticity accounting for higher-order interactions predicted the magnitude of subsequent change more accurately than pairwise-only plasticity, which substantially overestimated change when higher-order interactions were strong. This is consistent with the interpretation of local constraint as a prospective measure of the capacity for change, rather than a description of change already realized (\rf{branchi2026a}{Branchi, 2026a}).

A functional approach to higher-order interactions appears critical, because it captures aspects of a system that informational measures do not. This becomes evident when the property of interest is a functional capacity, such as plasticity. O-information was almost unrelated to the additional functional constraint, while information beyond pairs captured only part of its variation. The two quantities increased together when architecture was held constant, but diverged across different architectures. Thus, informational measures detect higher-order dependencies without necessarily quantifying how strongly those dependencies constrain change.

Informational indices have been related to the stability and flexibility of dynamics. In evolved Boolean networks, redundancy-dominated systems were more stable and synergy-dominated systems less stable (\rf{varley2024}{Varley and Bongard, 2024}). After ketamine, changes in redundancy have been proposed as a candidate marker of antidepressant-related neural reorganization (\rf{shah2026}{Shah et al., 2026}). The functional measure introduced here could complement these approaches by quantifying directly how strongly higher-order interactions constrain change.

These results build on existing network measures of plasticity. The local constraint shares with them the principle that plasticity is inversely related to realized constraint (\rf{dellicolli2024}{Delli Colli et al., 2024}; \rf{dellicolli2025}{Delli Colli et al., 2025}; \rf{viglione2025}{Viglione et al., 2025}; \rf{dellicolli2026}{Delli Colli et al., 2026}), and extends it to higher-order interactions. The extension matters most when pairwise couplings are weak and higher-order couplings strong. In this regime a system can appear loosely connected, and hence plastic, from its pairwise statistics while being substantially constrained. The same limitation applies to empirical measures that depend only on the means and simultaneous pairwise correlations of the variables. Whether measures based on pairs of trajectories, such as aggregated dynamic time warping, share this limitation was not tested here.

Two implications follow. First, a full account of plasticity requires higher-order constraints to be considered. Second, when the question concerns a functional property, such as plasticity, higher-order analysis should address the functional consequences of interactions, not only their informational signature. This is because the term higher-order does not denote a single property. The same structure has an informational and a functional reading, which can give different answers in the same systems, so studies should state which one they measure. In particular, an informational index such as O-information, whether it indicates synergy or redundancy, does not by itself indicate how strongly higher-order interactions constrain change. The functional and informational approaches therefore provide complementary descriptions of higher-order interactions.

\section{Limitations}

All quantities were computed exactly in simulated systems. The systems were small, so that every quantity could be computed exactly. Elements were binary and interactions stopped at the third order. The relation between plasticity and change was tested under a single dynamical rule, with the relation between the two estimated in pairwise-only systems. Estimating the higher-order constraint from empirical data, where the true couplings are unknown and observations are noisy, requires dedicated estimators and is left to future work.

\section{Conclusion}

Plasticity depends on constraints arising from both pairwise and higher-order interactions. These constraints are captured by a functional description of higher-order interactions rather than by informational indices, and accounting for them makes the prediction of change more accurate. Measuring realized constraint beyond pairwise interactions therefore provides a more complete characterization of the capacity of a system to change.

\section*{Acknowledgments}

I thank Claudia Delli Colli, Aurelia Viglione, Silvia Poggini for valuable discussion of the ideas presented in this paper. The research was supported by the grant from the ERANET Neuron and Istituto Superiore di Sanit\`a, project EnviroMood, and Istituto Superiore di Sanit\`a, project NeuroDynamics ISS20-6bbea6c1e02b, both to IB. The funders had no role in the preparation, review, or approval of the manuscript, or decision to submit the manuscript for publication.

\section*{Author Declarations}

I declare I have no competing interests.

The conception, the concepts and the theoretical argument presented here are entirely those of the author. Simulation and analysis code was written with the assistance of artificial intelligence assistants (Claude by Anthropic and ChatGPT by OpenAI, November 2026), which were also used to support the mathematical formalization. All code was checked and run by the author, who verified every derivation and result. I.B.: conceptualization, formal analysis, methodology, visualization, writing---original draft, writing---review and editing.

\section*{Data availability}

No empirical data were used in this study. The data that support the findings of this study were generated by simulation and are openly available, together with the code that generates them, in Zenodo at \url{https://doi.org/10.5281/zenodo.24561787}.

\section*{References}
\small
\begin{list}{}{\leftmargin=1.5em\itemindent=-1.5em\itemsep=0.3em\parsep=0pt}
\bibent{battiston2020}{Battiston F, Cencetti G, Iacopini I, Latora V, Lucas M, Patania A, Young J-G, Petri G (2020) Networks beyond pairwise interactions: Structure and dynamics. \textit{Physics Reports} 874:1--92. \doi{10.1016/j.physrep.2020.05.004}}
\bibent{branchi2022}{Branchi I (2022) Plasticity in mental health: A network theory. \textit{Neuroscience and Biobehavioral Reviews} 138:104691. \doi{10.1016/j.neubiorev.2022.104691}}
\bibent{branchi2023}{Branchi I (2023) A mathematical formula of plasticity: Measuring susceptibility to change in mental health and data science. \textit{Neuroscience and Biobehavioral Reviews} 152:105272. \doi{10.1016/j.neubiorev.2023.105272}}
\bibent{branchi2026a}{Branchi I (2026a) Quantifying plasticity: A network-based framework linking structure to dynamical regimes. \textit{Neuroscience and Biobehavioral Reviews} 187:106765. \doi{10.1016/j.neubiorev.2026.106765}}
\bibent{branchi2026b}{Branchi I (2026b) A theory of plasticity: capacity for change as inverse configurational constraint. arXiv:2609.25312. \doi{10.48550/arXiv.2609.25312}}
\bibent{dellicolli2025}{Delli Colli C, Viglione A, Giuliani A, Branchi I (2025) Anticipating depression trajectories by measuring plasticity and change through symptom network dynamics. \textit{European Psychiatry} 68:e128. \doi{10.1192/j.eurpsy.2025.10083}}
\bibent{dellicolli2026}{Delli Colli C, Viglione A, Poggini S, Branchi I (2026) Network-based assessment of plasticity predicts vulnerability to depressive symptomatology in healthy individuals. \textit{Journal of Affective Disorders} 412:122129. \doi{10.1016/j.jad.2026.122129}}
\bibent{dellicolli2024}{Delli Colli C, Chiarotti F, Campolongo P, Giuliani A, Branchi I (2024) Towards a network-based operationalization of plasticity for predicting the transition from depression to mental health. \textit{Nature Mental Health} 2:200--208. \doi{10.1038/s44220-023-00192-z}}
\bibent{glauber1963}{Glauber RJ (1963) Time-dependent statistics of the Ising model. \textit{Journal of Mathematical Physics} 4:294--307. \doi{10.1063/1.1703954}}
\bibent{jaynes1957}{Jaynes ET (1957) Information theory and statistical mechanics. \textit{Physical Review} 106:620--630. \doi{10.1103/PhysRev.106.620}}
\bibent{luppi2022}{Luppi AI, Mediano PAM, Rosas FE, Holland N, Fryer TD, O'Brien JT, Rowe JB, Menon DK, Bor D, Stamatakis EA (2022) A synergistic core for human brain evolution and cognition. \textit{Nature Neuroscience} 25:771--782. \doi{10.1038/s41593-022-01070-0}}
\bibent{rosas2019}{Rosas FE, Mediano PAM, Gastpar M, Jensen HJ (2019) Quantifying high-order interdependencies via multivariate extensions of the mutual information. \textit{Physical Review E} 100:032305. \doi{10.1103/PhysRevE.100.032305}}
\bibent{rosas2022}{Rosas FE, Mediano PAM, Luppi AI, Varley TF, Lizier JT, Stramaglia S, Jensen HJ, Marinazzo D (2022) Disentangling high-order mechanisms and high-order behaviours in complex systems. \textit{Nature Physics} 18:476--477. \doi{10.1038/s41567-022-01548-5}}
\bibent{schneidman2003}{Schneidman E, Still S, Berry MJ, Bialek W (2003) Network information and connected correlations. \textit{Physical Review Letters} 91:238701. \doi{10.1103/PhysRevLett.91.238701}}
\bibent{shah2026}{Shah K, Herzog R, Swann AC, O'Brien B, Balakrishnan R, Mathew SJ, Murphy N (2026) High-order brain interactions during ketamine-induced state changes: A functional marker of response in late-life treatment-resistant depression? \textit{Translational Psychiatry} 16:374. \doi{10.1038/s41398-026-04212-1}}
\bibent{varley2024}{Varley TF, Bongard J (2024) Evolving higher-order synergies reveals a trade-off between stability and information-integration capacity in complex systems. \textit{Chaos} 34. \doi{10.1063/5.0200425}}
\bibent{viglione2025}{Viglione A, Delli Colli C, Poggini S, Mobasher M, Vai B, Poletti S, Benedetti F, Branchi I (2025) A network-based approach reveals higher plasticity levels in bipolar than major depressive disorder. \textit{Journal of Affective Disorders} 387:119549. \doi{10.1016/j.jad.2025.119549}}
\end{list}

\end{document}